\documentclass[10pt,twocolumn,letterpaper]{article}

\usepackage{cvpr}
\usepackage{times}
\usepackage{epsfig}
\usepackage{graphicx}
\usepackage{amsmath}
\usepackage{amssymb}
\usepackage{comment}

\usepackage[pagebackref=true,breaklinks=true,letterpaper=true,colorlinks,bookmarks=false]{hyperref}
\DeclareMathOperator{\atantwo}{atan2}
\cvprfinalcopy 

\def\cvprPaperID{5} 

\ifcvprfinal\fi
\begin{document}

\title{Estimation of Orientation and Camera Parameters from Cryo-Electron Microscopy Images with Variational Autoencoders and Generative Adversarial Networks}

\author{Nina Miolane\\
Stanford University\\
{\tt\small nmiolane@stanford.edu}
\and
Fr\'ed\'eric Poitevin\\
Stanford University\\
{\tt\small frederic.poitevin@stanford.edu}
\and
Yee-Ting Li\\
SLAC National Accelerator\\
{\tt\small ytl@slac.stanford.edu}
\and
Susan Holmes\\
Stanford University\\
{\tt\small susan@stat.stanford.edu}
}

\maketitle
\thispagestyle{empty}
\begin{abstract}
    Cryo-electron microscopy (cryo-EM) is capable of producing reconstructed 3D images of biomolecules at near-atomic resolution. However, raw cryo-EM images are only highly corrupted - noisy and band-pass filtered - 2D projections of the target 3D biomolecules. Reconstructing the 3D molecular shape requires the estimation of the orientation of the biomolecule that has produced the given 2D image, and the estimation of camera parameters to correct for intensity defects. Current techniques performing these tasks are often computationally expensive, while the dataset sizes keep growing. There is a need for next-generation algorithms that preserve accuracy while improving speed and scalability. In this paper, we combine variational autoencoders (VAEs) and generative adversarial networks (GANs) to learn a low-dimensional latent representation of cryo-EM images. This analysis leads us to design an estimation method for orientation and camera parameters of single-particle cryo-EM images, which opens the door to faster cryo-EM biomolecule reconstruction.
    
\end{abstract}

\section{Introduction}

Cryo-electron microscopy (cryo-EM) is one of the most promising imaging techniques in biology, as it produces 3D reconstructions of biomolecules at near-atomic resolution. However, raw cryo-EM images are only highly corrupted 2D projections of the target 3D biomolecules. The required  reconstruction of a 3D molecular shape starts with the removal of image outliers, the estimation of the biomolecule's 3D orientation that has produced each 2D image, and the estimation of camera parameters at acquisition. This paper proposes a neural network architecture with geometric regularization in order to perform these three tasks in the latent space of a variational autoencoder.

\subsection{Cryo-EM 3D reconstruction}

Over the years, protocols for cryo-EM data collection and data processing have matured into a well-defined pipeline  \cite{Nogales2015TheTechnique, Fernandez-Leiro2017ARELION, Thompson2019CollectionMicroscopy}.  We assume the following image formation model under the weak phase approximation \cite{Frank1996Two-DimensionalTechniques}: 
\begin{equation}
    x_{i} = \text{PSF}_{i} * P[R_{i}.V] + \eta_{i}, \qquad i=1, .., n,
    \label{eq:formationmodel}
\end{equation}
where $x_i$ is a $p\times p$ particle image, $V$ is a 3D grid of $p\times p\times p$ voxels whose projection is convolved with the camera's point-spread function (PSF), and $\eta$ is the remaining part of the signal, typically assumed to be uncorrelated noise. The operator $P$ projects the 3D volume along the Z-axis, after the reference volume $V$ has been subjected to rotation $R_{i}$. We refer to $R_i$ as the ``orientation" of the particle. In general, the PSF is inherited from the micrograph from which the particle image was extracted, although it can also be further refined \emph{a posteriori} \cite{Grant2018CisTEMProcessing, Zivanov2018NewRELION-3}. We refer to the $\text{PSF}_i$ as the ``camera parameters" that have generated the particle. We see that reconstructing $V$ from the particle dataset amounts to estimating the rotation assuming they are all properly centered, and the PSF. Equivalently, this could be cast in a Bayesian formulation in a model with latent variables $\phi = (R,\text{PSF})$ \cite{Scheres2012ADetermination}.

In \cite{Scheres2012ADetermination}, the parameter V is learned through expectation maximization \cite{Dempster1977Algorithm}. It iteratively refines the estimate of $V$, that depend on the posterior probability of the orientation of each image, given the estimate of $V$ at the previous step. This posterior probability is calculated numerically by systematically evaluating the agreement between each image and all possible orientation of the current model, at each iteration. This costly step has greatly benefited from parallelization using GPUs and hybrid clusters \cite{Kimanius2016AcceleratedRELION-2} or the implementation of a branch-and-bound algorithm \cite{Punjani2017CryoSPARC:Determination} that rules out large regions of the search space, but it remains a computational bottleneck.

\subsection{Related work}

Deep learning techniques have started to penetrate the data processing pipeline, to speed-up traditional steps of cryo-EM processing and reconstruction \cite{Bendory2019Single-particleOpportunities}. Researchers have trained neural networks to perform fast automated particle picking \cite{Wang2016DeepPicker:Cryo-EM, Zhu2017AMicroscopy, Wagner2019SPHIRE-crYOLOCryo-EM, Bepler2018Positive-unlabeledMicrographs, Bepler2019ExplicitlySpatial-VAE}, particle pruning \cite{Sanchez-Garcia2018DeepMicroscopy}, as well as validation and estimation of the resolution of cryo-EM 3D reconstructions \cite{Avramov2019DeepMaps}. These advances have targeted and accelerated steps before or after the 3D reconstruction itself. They have not addressed the computational burden of the joint orientation estimations and volume reconstruction.

Recently, \cite{Zhong2019ReconstructingModels} used a spatial variational autoencoder (VAE) \cite{Bepler2019ExplicitlySpatial-VAE} to capture the heterogeneity of 3D structures within a continuous latent space. However, the architecture keeps a computational bottleneck in the orientation estimation required for volume reconstruction. The orientation $R$ is not estimated through the VAE, but with a global search using a branch and bound algorithm, similar to \cite{Punjani2017CryoSPARC:Determination}. In contrast, this paper specifically targets the orientation estimation using a VAE architecture.



Estimating the orientation implies being able to disentangle it from other variables in the latent space of a VAE. State-of-the-art methods for disentangling variables often rely on modifying the VAE loss function. The $\beta$-VAE heavily penalizes the regularization factor in the VAE loss and achieves good disentanglement results on images \cite{Thiagarajan2017-VAE:Framework}. However, it is unclear if this penalization works for all datasets. Other solutions choose to maximize the mutual information between a few latent variables and the observation (InfoGAN, \cite{Chen2016InfoGAN:Nets}), an approximation of the total correlation (TC) ($\beta$-TCVAE, \cite{Chen2018IsolatingAutoencoders}), or the Hilbert-Schmidt Independence Criterion (dHSIC) \cite{Lopez2018InformationBayes} to enforce independence between the latent representations and arbitrary nuisance factors. Similarly, the variational fair autoencoder (VFAE) \cite{Louizos2015TheAutoencoder} uses priors that encourage independence between sensitive and latent factors of variation, to learn invariant representation. 

Other solutions tackle disentanglement by encouraging a structure within the latent space of the VAE. FactorVAE \cite{Kim2018DisentanglingFactorising} enforces independence by encouraging a factorial decomposition of the latent space. The Variationally Inferred Transformational Autoencoder (VITAE) proposes an architecture with two latent spaces to separate spatial transformation from visual style such as shape \cite{Detlefsen2019ExplicitModels}. Likewise, the spatial VAE was designed to disentangle image orientation and translation from other latent variables \cite{Bepler2019ExplicitlySpatial-VAE}, by introducing a geometric structure in the decoder. In contrast, our approach disentangles the variables by exploiting the specific geometric structure of the cryo-EM data set.

\subsection{Contributions}

We approach disentanglement of orientation and camera parameters by studying the geometric properties of the cryo-EM images projected in the latent space of a combination of a variational autoencoder and a generative adversarial network. We show that the projected images possess a structure of ``orbits", in the sense of Lie group theory. Such a structure is expected given the acquisition procedure of cryo-EM images. We use this observation to design an estimation method that computes the orientation and camera parameters of a given image, after outliers removal.

\section{Elements of geometry}\label{sec:geom}

\subsection{Cryo-EM images and the action of rotations}

We consider 2D images with compact domain $\Omega \subset \mathbb{R}^2$. We adopt the point of view of images as square-integrable functions $x$ over the domain $\Omega$ ; i.e., we write $x \in L_2(\Omega )$, where $L_2(\Omega )$ is a Hilbert space. Cryo-EM images of a single particle, represented by a volume $V$ in $\mathbb{R}^3$, form a subspace of $L_2(\Omega)$. We first consider the ideal cryo-EM data space, written $\mathcal{C}_V(\Omega)$, which consists of hypothetical cryo-EM images acquired without noise and with constant camera parameters. This space is defined as:
\begin{align*}
&\mathcal{C}_V(\Omega) \\
 & = \{ x \in L_2(\Omega)\ |\ \exists ! R_x \in SO(3), x = \text{PSF} \ast P(R_x \cdot V)\},
\end{align*}
where $SO(3)$ is the group of 3D rotations, $R_x$ is the rotation corresponding of the particle volume's orientation at acquisition time, $P$ is a projection operator and $\text{PSF}$ summarizes the camera parameters. The uniqueness of $R_x$ allows to define the group action of $SO(3)$ below, but will not be required for the action of $SO(2)$.

To understand the geometry of the space $\mathcal{C}_V(\Omega)$, we define the following action of the 3D rotations on cryo-EM images: 
\begin{align*}
\rho : SO(3) \times \mathcal{C}_V(\Omega) &\rightarrow \mathcal{C}_V(\Omega),\\
(R , x) &\rightarrow R \cdot x = \text{PSF} \ast P(R \cdot R_x \cdot V).
\end{align*}
Mathematically, $\rho$ is called a Lie group action of the Lie group $SO(3)$ on the space $\mathcal{C}_V(\Omega)$. We refer to \cite{Postnikov2001, Alekseevsky2003, Huckemann2010IntrinsicActions} for details on Lie group actions.

Considering cryo-EM particles obtained from a single volume $V$ with inplane rotations, we can define the subspace $\mathcal{C}_V^{2D}(\Omega)$ of $\mathcal{C}_V(\Omega)$ as:
\begin{align*}
&\mathcal{C}_V^{2D}(\Omega) \\
 & = \{ x \in L_2(\Omega)\ |\ \exists R_x \in SO(2), x = \text{PSF} \ast P(R_x \cdot V)\},
\end{align*}
where the axis of the rotation in $SO(2)$ corresponds to the axis of the projection $P$. The action $\rho$ of the subgroup $SO(2)$ on the subspace $\mathcal{C}_V^{2D}(\Omega)$ is a restriction to $\mathcal{C}_V^{2D}(\Omega) $ of the action $\rho'$ of $SO(2)$ on 2D images, defined as follows:
\begin{align*}
\rho' : SO(2) \times L_2(\Omega) &\rightarrow L_2(\Omega),\\
(R , x) &\rightarrow R \cdot x = x \circ R^{-1},
\end{align*}
where the image's domain is rotated by $R^{-1}$. From now on, we restrict our focus to the cryo-EM image subspace $\mathcal{C}_V^{2D}(\Omega)$ within the image space $L_2(\Omega)$ equipped with the action $\rho'$ of 2D rotations.

\begin{figure}[h!]
\centering
\def\svgwidth{0.6\columnwidth}
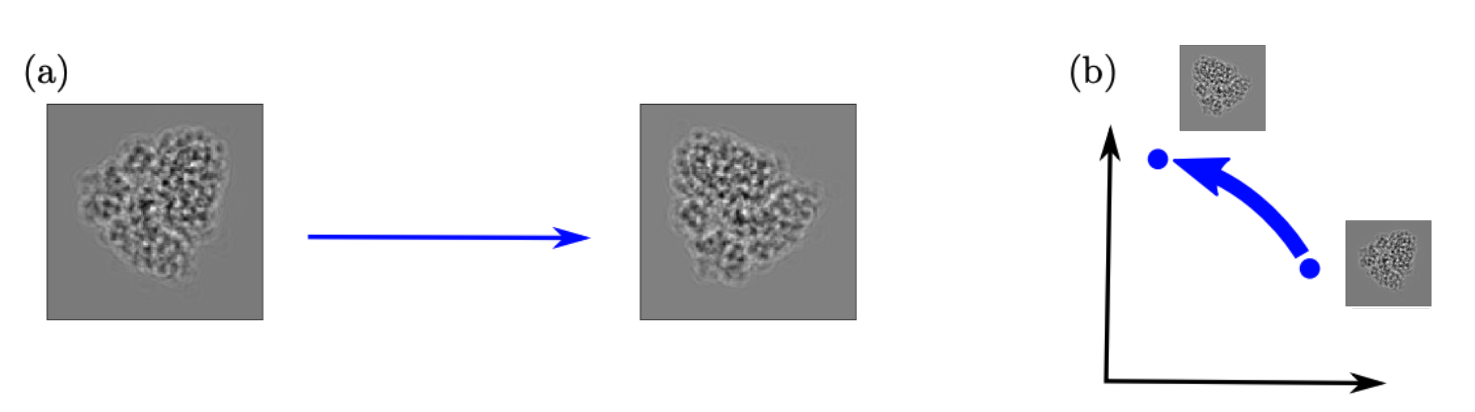
    \caption{Action of 2D rotations on the space of images $L_2(\Omega)$, schematically represented as $\mathbb{R}^2$.}
    \label{fig:cryo_math}
\end{figure}

Figure~\ref{fig:cryo_math} represents the space of images $L_2(\Omega)$ as the toy Hilbert space $\mathbb{R}^2$. It shows two cryo-EM images $x_1, x_2 \in \mathcal{C}_V^{2D}(\Omega)$ from the same particle $V$, and another image $x_3$, as three points in this space. The action of $SO(3)$ is schematically represented there: the image $x_1$ is transformed into the image $x_2 = R \cdot x_1$ by the action of the rotation $R$.

\subsection{Orbit and isotropy group}

The orbit $O_x$ of an image $x \in L_2(\Omega)$ by the action $\rho'$ of $SO(2)$ is defined as the images reachable through the action of rotations on $x$: $O_x = \{ x' \in L_2(\Omega)\ |\ \exists R \in SO(2), x' = R \cdot x\}.$ On Figure~\ref{fig:cryo_math}, the orbit $O_{x_1}$ of $x_1$ is represented as the green dotted circle. From the definition, $O_{x_1}$ is also the orbit of $x_2$ and the two images $x_1$, $x_2$ are points on the green dotted circle. 

The isotropy group $G_x$ of a image $x$ by the action $\rho'$ is defined as the subgroup of $SO(2)$ formed by the rotations that leave $x$ unchanged: $G_x = \{ R\in SO(2)\ |\  R \cdot x = x\}$. $G_x$ describes the intrinsic symmetry of the image $x$ and the volume $V$: the more symmetric is $x$, the larger its isotropy group, where large is understood in the sense of inclusion. 

The isotropy group of the images $x_1, x_2$ in Figure~\ref{fig:cryo_math} is restricted to the identity $\mathbb{I}_2$ of the group $SO(2)$. Only the identity leaves the images $x_1$ and $x_2$ unchanged. In contrast, the isotropy group of the image $x_3$ is the whole Lie group of 2D rotations $SO(2)$. Any rotation leaves this image invariant. The image $x_3$ shows more symmetry than $x_1$ and $x_2$, and thus has a larger isotropy group. The asymmetric details of $x_1$, $x_2$ are the sign of a smaller isotropy group. 

As an intermediate example, consider an image exhibiting $C_n$ symmetry, where $C_n$ is the group of $n$-fold rotations corresponding to the set of $2D$ rotations with angles in $\frac{2\pi}{n}$ $\mathbb{Z} / n\mathbb{Z}$. Such image has a greater isotropy group than $x_{1}$ and $x_{2}$ since $n-1$ rotations leave it unchanged, on top of the identity $\mathbb{I}_2$. But it exhibits a smaller isotropy group than $x_{3}$ since many rotations do affect it.

\subsection{Ideal cryo-EM data space $\mathcal{C}^{2D}_V(\Omega)$}

We describe the geometry of the ideal cryo-EM data space with in-plane rotations $\mathcal{C}^{2D}_V(\Omega)$, \textit{i.e.} cryo-EM images acquired without noise and with constant camera parameters, using the orbit-stabilizer theorem. This theorem links the isotropy group (also called the stabilizer) and the orbit of an image.

\begin{theorem}[Orbit-stabilizer Theorem]\label{th:orbitstab}
Consider a Lie group $G$ acting on a space $H$. Let $x\in H$ be an element of this space. Then, the orbit and isotropy group of $x$ are related by: $O_x \sim G / G_x$ where $\sim$ denotes an isomorphism and $/$ the quotient of groups.
\end{theorem}

Theorem~\ref{th:orbitstab} states that an image with a large isotropy group - \textit{i.e.} an image with a lot of symmetries - has a smaller orbit.  On Figure~\ref{fig:cryo_math}, the images $x_1$, $x_2$ have a small isotropy group and a circular orbit. In contrast, the image $x_3$ has a larger isotropy group and its orbit is a single point: $x_3$ itself.

Using the definition of the space $\mathcal{C}^{2D}_V(\Omega)$, we have the following result.

\begin{proposition}\label{prop}
The space of cryo-EM images $\mathcal{C}^{2D}_V(\Omega)$ is isomorphic to  $SO(2) / G_x$:
\begin{equation}
    \mathcal{C}^{2D}_V(\Omega) \sim SO(2) / G_x,
\end{equation}
where $G_x$ is the isotropy group of any image under the action $\rho'$.
\end{proposition}
The symmetries in the slice of volume $V$ considered dictates the geometry of the space of the 2D cryo-EM images with inplane rotations.

\subsection{Observed cryo-EM data space}

We extend the discussion to the space of observed cryo-EM with in-plane rotations, \textit{i.e.}, taking into account possible variations of the PSF and the noise model. The actual space of cryo-EM images is obtained by perturbing the space $\mathcal{C}^{2D}_V(\Omega)$ and can be interpreted as a ``noisy" version of $\mathcal{C}^{2D}_V(\Omega)$, with a complex model of noise. 

Locally at each cryo-EM image $x\in \mathcal{C}^{2D}_V(\Omega)$, the space of 2D images decomposes as a sum: $L_2(\Omega) = T_x\mathcal{C}^{2D}_V(\Omega) \oplus V_x$, where $T_x\mathcal{C}^{2D}_V(\Omega)=T_xO_x$ is the tangent space of the orbit $O_x$ and $V_x$ is a supplementary space. The space $V_x$ represents the variations in images that do not correspond to 2D rotations, for example, the effects of camera parameters on the image intensity. Our objective is to learn a latent representation of the space of images where we can fit the space $\mathcal{C}^{2D}_V(\Omega)$ by fitting a spherical subspace to our projected data. Then, decomposing the latent space of images as $T_x\mathcal{C}^{2D}_V(\Omega) \oplus V_x$, we use the coordinate on $T_x\mathcal{C}^{2D}_V(\Omega)$ to describe the image's orientation, and the coordinate on $V_x$ to describe the other parameters.

\section{Methods}\label{sec:meth}


\subsection{VAE-GAN architecture and training}

\subsubsection{Review of VAE}

We detail the combination of VAE \cite{Kingma2014Auto-EncodingBayes, Rezende2014StochasticModels} and GAN \cite{Goodfellow2014GenerativeNets} used in this paper. Consider a dataset $x_1, ..., x_n \in \mathbb{R}^D$. A VAE models each data point $x_i$ as the realization of a random variables $X_i$ generated from a nonlinear probabilistic model with lower-dimensional unobserved latent variable $Z_i$ taking value in $\mathbb{R}^L$, where $L < D$, such as:
\begin{equation}\label{eq:vae}
    X_i = f_\theta(Z_i) + \epsilon_i,
\end{equation}
where $\epsilon_i$ represents \textit{i.i.d.} measurement noise and follows a multivariate isotropic Gaussian: $\epsilon_i \sim N(0, \sigma^2 \mathbb{I}_D)$, and $Z_i  \sim N(0, \mathbb{I}_L)$. The function $f_\theta$ belongs to a family $\mathcal{F}$ of nonlinear generative model parameterized by $\theta$, and is typically represented by a neural network, called the decoder.

The VAE fits model~(\ref{eq:vae}) by training the decoder, while also training an encoder that simultaneously learns a distribution $q_\phi(z|x)$, within a variational family $\qdists$ parameterized by $\phi$, of the posterior distribution of the latent variables $Z_i$. The VAE achieves its objective by minimizing an upper-bound of the negative log-likelihood, which writes as the sum of a reconstruction and regularization losses:
\begin{align*}
    \mathcal{L}_{\text{VAE}}
    &= \mathcal{L}_{\text{rec}} + \mathcal{L}_{reg}\\
    &= -\mathbb{E}_{q_\phi(z)} \left[ \log p_\theta(x|z) \right] + \text{KL} \left( q_\phi(z|x) \parallel \prior \right),
\end{align*}
where KL is the Kullback-Leibler divergence.

From a geometric perspective, VAE performs manifold learning as it learns $\theta$ to fit the manifold $f_\theta(\mathbb{R}^L)$ to the data. The function $f_\theta$ being continuous, VAEs can only learn manifolds homeomorphic to $\mathbb{R}^L$. As a consequence, a VAE needs a least two latent dimensions to capture a circle. For the simulated dataset with in-plane rotations, we choose $L=3$ to capture a circle and an additional dimension of defocus. For the experimental dataset with in-plane rotations, we choose $L=4$ to capture a circle, a dimension of defocus, and we add a dimension to capture other parameters like nuisance variables.

\subsubsection{Review of GAN}

We enhance the VAE by adding a GAN \cite{Boesen2016AutoencodingMetric}. A GAN is a method to fit a model using an adversarial process. The GAN jointly fits the generative model to best capture the data distribution, and a discriminative model $D$ that estimates the probability that a $x_i$ came from the training data rather than the generative model.

The GAN achieves its objective by finding the binary classifier that gives the best discrimination between true and generated data and simultaneously encouraging the generator to fit the true data distribution. In practice, a GAN maximizes/minimizes the binary cross-entropy:
\begin{align*}
    \mathcal{L}_{\text{GAN}} = \log p_{\text{Dis}}(x) + \log \left( 1- p_{\text{Dis}} (\text{Gen}(z)) \right)
\end{align*}
with respect to the discriminator or generator with $x$ being a training sample.

From a geometric perspective, a GAN implicitly has to learn a complex similarity metric to discriminate actual data from generated data. For images, GAN usually produces reconstruction images of better visual quality. Learning an implicit metric is particularly interesting for cryo-EM images that are highly corrupted by noise.

\subsubsection{VAE-GAN with geometric regularization}

\begin{figure}[h!]
\centering
\def\svgwidth{0.9\columnwidth}
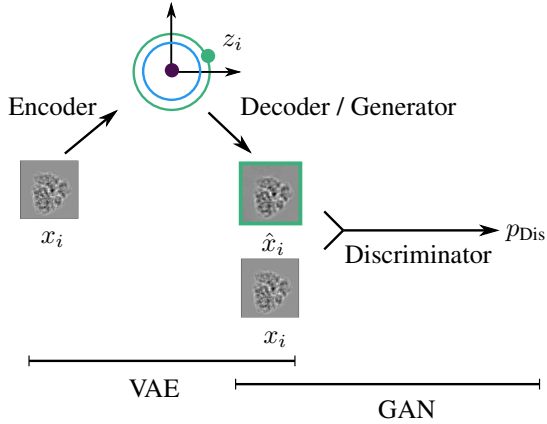
    \caption{VAE-GAN architecture, illustrated with the simulated images. The encoder encodes the image $x_i$ into a latent variable $z_i$, that is decoded into a reconstructed image $\hat{x}_i$ by the decoder. The discriminator estimates the probability that $x_i$, $\hat{x}_i$ are real images.}
    \label{fig:cryo_vaegan}
\end{figure}

We add the GAN as a refinement to the reconstruction loss of the VAE. Our final architecture is shown on Figure~\ref{fig:cryo_vaegan}. It differs from \cite{Boesen2016AutoencodingMetric} in two ways: (i) we keep the VAE's reconstruction term, implemented with a binary cross-entropy, (ii) we do not use the $L_2$ norm on the discriminator features. Finally, we add a geometric regularizer to encourage the latent variables to form a cone: $\mathcal{L}_{\text{cone}} = (z_1^2 + z_2^2 - z_3^2)^2$, where $z_1, z_2, z_3$ are the first three dimensions of the latent space. Our loss writes:
\begin{align*}
    \mathcal{L}_{\text{VAEGAN}} 
    = \mathcal{L}_{\text{rec}} 
    + \lambda_{\text{reg}} \mathcal{L}_{\text{reg}}
    + \lambda_{\text{GAN}} \mathcal{L}_{\text{GAN}}
    + \lambda_{\text{cone}} \mathcal{L}_{\text{cone}}
\end{align*}
where $\lambda_{\text{reg}}, \lambda_{\text{GAN}}, \lambda_{\text{cone}}$ are hyperparameters.

The implementation of the encoder consists of convolutional layers with batch normalization \cite{Ioffe2015BatchShift} and a fully connected layer to predict the parameters of the variational distribution. The implementation of the decoder consists of convolutional layers, padding, and batch normalization. 

\subsubsection{Model selection with hyper-parameter search}

During training, the model is initialized with the standard heuristic \cite{Glorot2010UnderstandingNetworks}. For optimization, we use Adam \cite{Kingma2015Adam:Optimization} with standard parameters $\beta_1 = 0.9$, $\beta_2 = 0.999$. We use mini batches of size $B=16$. 

We tune the learning rate, the number of decoder's layers, and the hyperparameters $\lambda_{\text{reg}}, \lambda_{\text{GAN}}$ with Bayesian optimization using the Hyperopt algorithm \cite{Bergstra2012RandomOptimization} and the Asynchronous Successive Halving Algorithm as scheduler \cite{Li2018MassivelyTuning} optimizing the validation loss. We do not tune hyperparameter $\lambda_{\text{cone}}$ which is set to $1$ as to represent our geometric constraint.

\subsection{Exploratory Analysis and Estimations}

After training of VAE-GAN, we project the cryo-EM dataset in the learned latent space. Section~\ref{sec:geom} established that the latent space decomposes, at a given cryo-EM image $x$, as $T_xO_x \oplus V_x$  where $O_x$ is the orbit of $x$ and $V_x$ is a supplementary space. Taking $V_x$ orthogonal to the orbit, we capture variations in the image that are rotationally invariant (\emph{e.g.} defocus or other camera parameters). In this subsection, we curate and interpret the latent space in terms of those subspaces.

\subsubsection{Outliers detection}

We curate the dataset projected in the latent space by removing outliers. We expect most of the dataset to correspond to non-outliers images that come from the generative model of Equation~\ref{eq:formationmodel} and share strong similarities. As such, the projections of these images are clustered together tightly in the latent space. In contrast, outliers are not drawn from the generative model and are likely to differ substantially. Their projections locate themselves at a distance from the non-outliers data. 

We use and compare outlier detection methods for the data projected in the latent space: robust covariance (RobCov) method \cite{Gnanadesikan2014RobustData}, isolation forest (IF) method \cite{Liu2008IsolationForest:Forest} and local outlier factor (LOF) methods \cite{Jahanbegloo2012IdentifyingOutliers}. They all assign a measure $m_{i}$ to each data point $i$, and a threshold $m_{max}$ value that can be used to assign outlier status to a point when $m_{i} > m_{max}$. For RobCov, the measure $m$ is the Mahalanobis distance centered at the origin of the latent space and with parameters given by the Hessian of the data. For IF, the data is represented as a tree through recursive partitioning, and the measure $m$ is related to the path length averaged over a forest of random trees. For LOF, the measure $m$ is related to the local density of the data around each data point. 

We use the output of RELION to attribute the ground-truth label outlier/non-outlier to each datapoint \cite{Scheres2015Semi-automatedRELION-1.3}. Precisely, we consider the Z-score computed for each image in RELION, either at the 2D classification or 3D refinement steps, depending on the dataset considered, see Section~\ref{sec:data}. A data point is an outlier if its Z-score is larger than 2 in absolute value, it is a non-outlier otherwise. To assess the diagnostic ability of each outlier detection method, we measure the area under the curve (AUC) of the receiver operating characteristic (ROC) curve that is created by plotting the true positive rate (TPR) against the false positive rate (FPR) as $m_{max}$ is varied for each method. 

\subsubsection{Estimation of defocus and orientation}


We carry out principal component analysis (PCA) to identify the three dimensions bearing most of the variability: $\mathbf{Z} = \mathbf{U}\mathbf{\Sigma}\mathbf{V}^{T}$, where $Z$ is the design matrix of the data in the latent space. We call ``reduced space" the space formed by the first two principal components of the PCA. The projection of the dataset in the reduced space is stored in the matrix $\mathbf{U}\Sigma$. We use the reduced coordinates $\mathbf{U} = (U_{1}, U_{2})$ to yield defocus and orientation by converting them into polar coordinates:
\begin{equation}
    (U_{1,i}, U_{2,i}) \rightarrow \big(r_{i} = \sqrt{U_{1,i}^{2} +  U_{2,i}^{2}}, \theta_{i} = \atantwo{\frac{U_{2,i}}{U_{1,i}}}\big).
\end{equation}
We estimate the defocus of image $i$ with the formula:
\begin{equation}
    \widehat{\text{def}}_i = (q_3 - q_1) \frac{r_i - q_1(r)}{q_3(r) - q_1(r)} + q_1,
\end{equation}
where $q_1(r), q_3(r)$ are the first and third quartiles of the empirical distribution of radii, and $q_1, q_3$ are the corresponding quartiles of the uniform distribution on the known range of defocuses $[0.5\mu\text{m}, 2.5\mu\text{m}]$. We estimate the orientation of image $i$ as the angle $\theta_i$ given by the polar coordinates in the reduced space. 

We evaluate the accuracy of this estimation procedure using the mean square error (MSE) between our predictions and the corresponding ground-truth, which is taken to be the defocus and orientation assigned by RELION. As we cannot expect to have the same origin of orientations nor the same orientation direction as RELION, we authorize a global shift $\Delta$ and a global change of sign when evaluating the MSE in orientations. 

Furthermore, keeping in mind the cryo-EM downstream task that uses orientation estimates to reconstruct the 3D molecular shape, we allow the possibility of down-weighting each estimate, based on their assumed quality or precision. At reconstruction time, we can down-weight the associated images accordingly. We use the estimated defocus to build the weight as $w_{i} = e^{\widehat{\text{def}}_i} - 1$. The weight $w_{i}$ reflects the level of confidence in the corresponding orientation estimate for image $i$: we are less confident in the orientation estimate corresponding to images with high defocus, \textit{i.e.} blurry images. 

Finally, the MSE formulas used to evaluate the accuracy of the defocus and orientation estimates write:

\begin{align}
    \text{MSE}_{\theta} &= \text{argmin}_{\Delta, \pm } \big(\sum_{i}^{n}w_{i}(\pm\theta_{i} + \Delta - \theta_{i}^{\text{(true)}} )^{2}\big) \\\nonumber
    \text{MSE}_{\text{def}} &= \frac{1}{n}\sum_{i}^{n}(\widehat{\text{def}}_{i}  - \text{def}_{i}^{\text{(true)}} )^{2}.
\end{align}

\section{Datasets}\label{sec:data}

\subsection{Simulated dataset}

We first created an ideal simulated dataset for which the simulation of micrographs was carried out using a transmission electron microscopy (TEM) simulator \cite{Rullgard2011SimulationSpecimens}. The TEM simulator simulates the process of cryo-EM images acquisition. We detail below the values of the parameters used required by the simulator. We used a single structural model of the human 80S ribosome \cite{Anger2013StructuresRibosome}. We chose disks of diameter 1200 nm for the simulated grid holes, with an even ice thickness of 100 nm. We set the acceleration voltage of the simulated electron beam at 300 kV, with an energy spread of 1.3 V. We set the electron dose per image to 100 e/nm$^{2}$. 

In terms of image acquisition parameters, we set the magnification to 81000 with spherical and chromatic aberrations set at 2.7 mm. The aperture diameter was set at 50 $\mu$m, the aperture angle at 0.1 mrad, and the focal length at 3.5 mm. The detector was defined as an array of 5760x4092 pixels, each of physical size 5 $\mu$m, thus setting the pixel size in the images to 0.62$\AA$. The simulation of noise was turned off, and the detector transfer function was assumed to be perfect. Micrographs were simulated assuming no motion of the particles, that were precisely placed in the field of view, while randomly rotated around the axis perpendicular to the grid plane (in-plane rotation). Micrographs were generated for each of the defocus values in the series going from 0.5 to 3.0 $\mu$m in steps of 0.5 $\mu$m. Each micrograph contains 48 images of the ribosome particle, extracted with size 648x648.

After this processing, we get a simulated dataset of $n=2,544$ cryo-EM images that we downsample to size 128x128. In this dataset, (i) there is one biomolecule's shape: the chosen structural model of the human ribosome 80s, (ii), this molecule is rotated in a plane, (iii) the camera parameters are restricted to a set of 6 different defocuses by design, (iv) there is no noise. Images randomly sampled from this dataset are shown on the first line in Figure~\ref{fig:cryo_recon_sim}.

\subsection{Experimental datasets}

We added experimental datasets created from samples of human 80S ribosomes. These samples were prepared and imaged, yielding 7664 micrograph movies. We use RELION 3.0 \cite{Zivanov2018NewRELION-3} to process the micrographs. Preprocessing steps involved motion correction \cite{Zheng2017MotionCor2:Microscopy} and CTF estimation \cite{Rohou2015CTFFIND4:Micrographs}. Particle picking was carried out using RELION's ``Autopick'' tool, yielding more than one million particles extracted as images rescaled to 180x180 with pixels of size 2.052 \AA. 

The resulting set of images was subjected to three successive rounds of 2D classification in RELION, yielding a final dataset of 279,261 particles. We have extracted class 93 from the second round of 2D classification, as well as classes 30 and 39 from the third round. We compute the z-scores assigned to each image by running RELION's ``Particle sorting"  tool on the respective 2D classification jobs: the z-score is a confidence score in the attribution of the given 2D class to the corresponding data point.


As a result of this processing, we have three experimental datasets of cryo-EM images, that we downsampled to size 128x128. In each of these datasets, (i) there is one biomolecule shape: the human 80S ribosome, (ii), this molecule is rotated in a plane. The first experimental inplane dataset (view 39) represents the real dataset equivalent to the simulated case and has $n = 5,119$ images. The second (view 93) has $n = 8,278$ images. The third (view 30) has $n = 4,917$ images.

\section{Results}\label{sec:res}

\begin{table}[]
\centering
\begin{tabular}{llll}
\hline
\multicolumn{1}{|l|}{}            & \multicolumn{1}{l|}{AUC } & \multicolumn{1}{l|}{Angle (degr.)}              & \multicolumn{1}{l|}{Defocus ($\mu$m)}     \\ \hline
\multicolumn{1}{|l|}{Sim.}         & \multicolumn{1}{l|}{N/A}          & \multicolumn{1}{l|}{1.72}            & \multicolumn{1}{l|}{0.18}        \\ \hline
\multicolumn{1}{|l|}{Exp. (view 93)}    & \multicolumn{1}{l|}{0.79}         & \multicolumn{1}{l|}{37.38 }                      & \multicolumn{1}{l|}{0.44}        \\ \hline
\multicolumn{1}{|l|}{Exp. (view 39)}    & \multicolumn{1}{l|}{0.86}         & \multicolumn{1}{l|}{18.61 } & \multicolumn{1}{l|}{0.45} \\ \hline
\\
\end{tabular}

\caption{AUC and weighted rMSE for angle and defocus estimations, for the simulated dataset and the experimental datasets of views 93 and 39. AUC values are given for (isolation forest; Z=2).}
\label{tab:estimations}
\end{table}

\subsection{Training VAE-GAN learns successive orbits}

\begin{figure}[h!]
    \includegraphics[width=8.3cm]{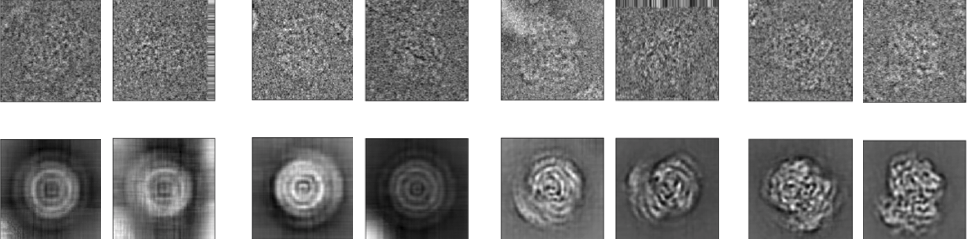}
    \caption{Reconstructions from simulated dataset (first two lines) and one experimental dataset (view 93, next two lines) during training, at four different epochs corresponding to the four columns.}
    \label{fig:cryo_recon_sim}
\end{figure}

We train VAE-GAN on the simulated dataset and the three experimental datasets. Figure~\ref{fig:cryo_recon_sim} shows the reconstructed images at different epochs during training, for the simulated dataset and the experimental dataset with view 93. 

We observe that VAE-GAN successively learns the different Fourier components of the images. From a geometric perspective, VAE-GAN starts with a prior distribution on the latent variables that concentrates at the origin of the latent space, which represents the most symmetric image. During training, VAE-GAN expands this prior distribution to populate the different orbits of the action of the rotations. Starting from the origin of the latent space, it grows circles of increasing radii. Therefore, rotational symmetries successively disappear, as images find their place on their respective orbits. 

At the end of the training, the images reconstructed from the simulated dataset are visually almost indistinguishable from the original images, see Figure~\ref{fig:cryo_recon_sim}. Each low-dimensional latent variable has captured the information specific to its image, while the decoder's parameters have captured what is invariant across the images: the shape of the ribosome. The images reconstructed from the experimental dataset show a molecular shape that we recognize as the ribosome, see Figure~\ref{fig:cryo_recon_sim}. The VAE-GAN has performed denoising, an indication that we have captured the space $\mathcal{C}_V^{2D}(\Omega)$ introduced in Section~\ref{sec:geom}.

\begin{figure}[h!]
    \centering
    \includegraphics[width=8.3cm]{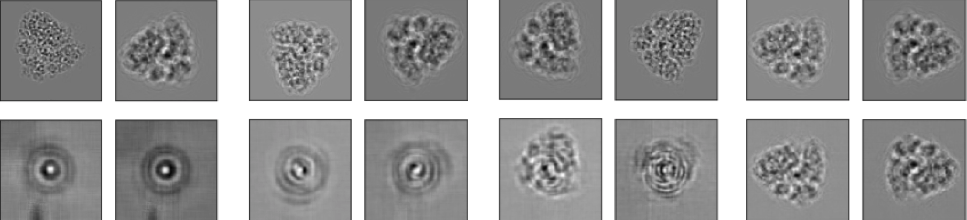}
\end{figure}

\subsection{Latent space of the simulated dataset}

\begin{figure}[h!]
\centering
\def\svgwidth{1\columnwidth}
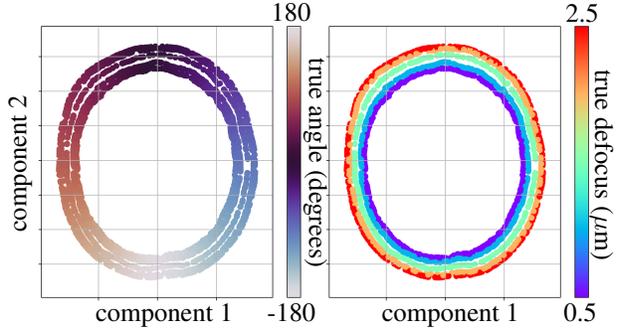
    \caption{Cryo-EM images from the simulated dataset projected in thelatent space. Left: colored by the angle of the in-plane rotation. Right: colored by defocus. }
    \label{fig:cryo_latent_sim_mus}
\end{figure}

We perform an exploratory analysis in the latent space of the simulated dataset. Figure~\ref{fig:cryo_latent_sim_mus} shows the data projected on the first two components of the principal component analysis, colored by defocus and orientation's angle. We observe that the latent space has learned the structure of orbits from the original space of images $L_2(\Omega)$, and has disentangled the camera parameter (defocus) from the ribosome's orientation. In polar coordinates, the defocus is related to the radius, while the ribosome's orientation is given by the angle, as shown in Figure~\ref{fig:cryo_latent_sim_mus}. We had six values of defocuses in the simulated dataset, yet we observe that the last three have not been satisfactorily clustered. The defocus levels had made the corresponding images indeed not distinguishable at the resolution chosen after downsampling.

\subsection{Latent space of inplane experimental datasets}

We perform an exploratory analysis in the latent spaces of the experimental datasets with in-plane rotations.

\subsubsection{Outlier detection}

\begin{figure}[h!]
\centering
\def\svgwidth{\columnwidth}
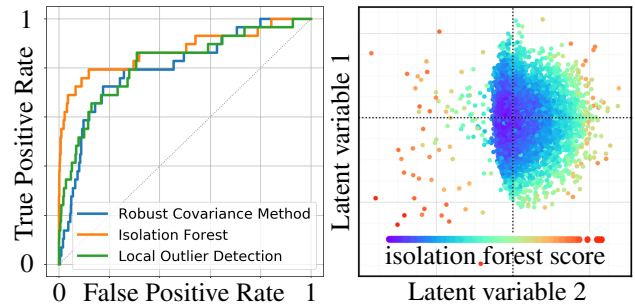
    \caption{Left: ROC curve for outlier detections in the latent space. We use a threshold $z=2$ from the ``ground truth" given by the traditional computationally expensive methods. Right: Outlier detection in the latent space with isolation forests. }
    \label{fig:roc_39}
\end{figure}

Figures~\ref{fig:roc_39} shows the data projected in the latent space of the experimental datasets with view 39, which is the view corresponding to the simulated dataset. We observe the same patterns as in the simulated case, with the addition of outliers that correspond to the corrupted images. The ROC curve of Figure~\ref{fig:roc_39} shows the results of the outliers detection methods, which efficiently reveal the outliers. For example, we report a AUC of 0.86 for the isolation forest method, see Table~\ref{tab:estimations}. The outlier detection methods perform equivalently well for the other experimental datasets, as shown in Table~\ref{tab:estimations}. 

\subsubsection{Orbits in the absence of symmetry}

\begin{figure}[h!]
\centering
\def\svgwidth{\columnwidth}
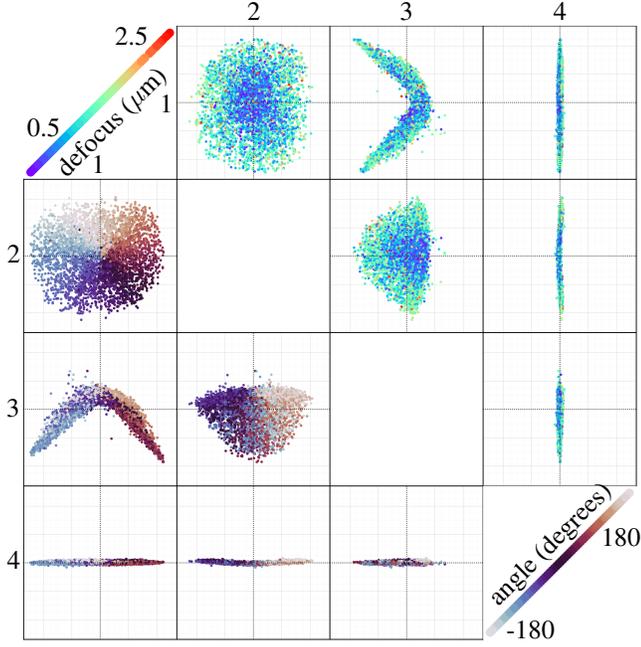
    \caption{Cryo-EM images from the experimental dataset on view 39 in latent space (L=4).}
    \label{fig:latent_39}
\end{figure}

We perform an exploratory analysis in the latent space of the first two experimental datasets (views 39 and 93), for which the 2D projection of the ribosome shape has no apparent symmetry. The numerical results are presented in Table~\ref{tab:estimations}. We show the principal components of the latent space for view 39 in Figure~\ref{fig:latent_39}. 

As the images in view 39 do not show any obvious symmetry, each orbit is isomorphic to $S^1$ as per Proposition~\ref{prop}. We observe a stratification of the latent space in concentric circles. As in the simulated case, the geometry of the group action leads to a natural disentanglement of the camera parameter (defocus) from the orientation's angle, which can be extracted by polar coordinates, see Figure~\ref{fig:latent_39}. Table~\ref{tab:estimations} summarizes corresponding rMSEs. We record similar rMSEs for the view 93, see Figure~\ref{fig:preds} (B). These results indicate that (unsupervised) estimation of defocus and orientations are possible using the latent space learned by VAE-GAN.

\subsubsection{Orbits in the presence of symmetry}

\begin{figure}[h!]
\centering
\def\svgwidth{0.8\columnwidth}
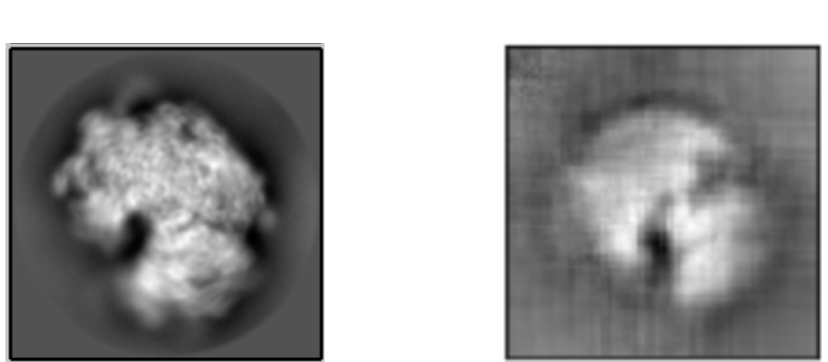
    \caption{Symmetries of the particle in view 30. (a) 2D class average taken from RELION. (b) VAE-GAN reconstruction, suggesting a $C_{3}$ symmetry.}
    \label{fig:cryo_symmetries}
\end{figure}

We perform an exploratory analysis in the latent space of the third experimental datasets (view 30). In this dataset, the 2D projection of the ribosome's shape exhibits symmetry, at the scale of the quality of the image reconstructions by the decoder, see Figure~\ref{fig:cryo_symmetries} (b). The symmetry can be described by the cycle group of 3-fold rotations, written $C_3$, which consists of the set of 2D rotations of angles $0, 2\pi/3$, and $4\pi/3$. As a consequence, each orbit is isomorphic to $S^1 / C_3$ as per Proposition~\ref{prop}.

\begin{figure}[h!]
\centering
\def\svgwidth{0.9\columnwidth}
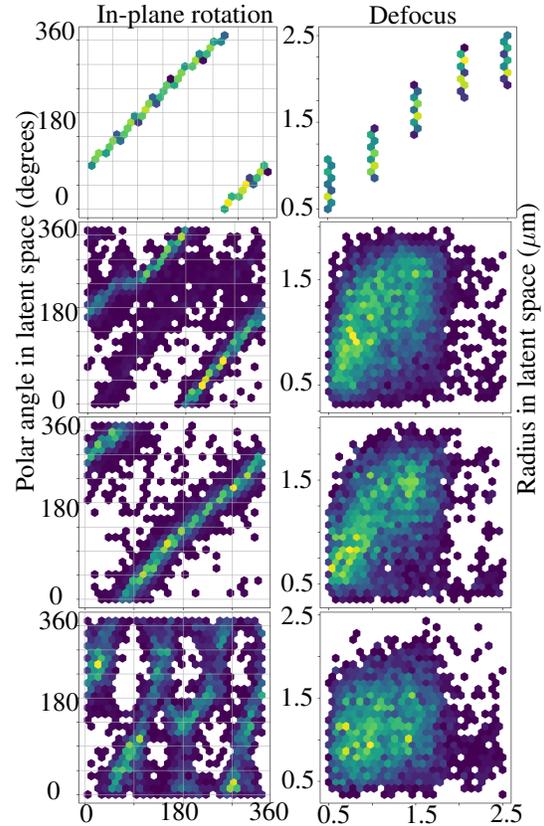
    \caption{Unsupervised estimations of angles (left) and defocus (right), compared to Relion (x-axis). \textbf{(top)} 
    Simulated dataset. Hexagonal bins with at least one data point are plotted. Highly populated bins are brighter. 
    \textbf{(bottom three)} Experimental views 93,39,30 going from top to bottom.}
    \label{fig:preds}
\end{figure}

We observe that the latent space has captured this geometry, which is revealed in Figure~\ref{fig:preds} (D, left) by the periodicity on the horizontal axis. A polar angle $\alpha$ in the latent space maps to $\alpha$, or $\alpha + 2\pi/3$, or $\alpha + 4\pi/3$ in terms of the ``true" angle. Additionally, we observe symmetry in reflection, corresponding to the symmetry on the vertical axis. From a mathematical perspective, it is remarkable that VAE-GAN has captured the geometry of the orbit $S^1 / C_3$.


\section{Conclusion}

We have combined variational autoencoders (VAEs) and generative adversarial networks (GANs) to learn the latent space of cryo-EM images. This new avenue to cryo-EM reconstruction would potentially unlock issues in scalabilities that traditional methods are starting to encounter in front of the fast-growing rate of data collection.

{\small
\bibliographystyle{ieee_fullname}

\begin{thebibliography}{10}\itemsep=-1pt

\bibitem{Alekseevsky2003}
Dmitri Alekseevsky, Andreas Kriegl, Mark Losik, and Peter~W Michor.
\newblock {The Riemannian Geometry of Orbit Spaces. The Metric, Geodesics, and
  Integrable Systems}.
\newblock {\em Publ. Math. Debrecen}, 62, 2003.

\bibitem{Anger2013StructuresRibosome}
Andreas~M. Anger, Jean~Paul Armache, Otto Berninghausen, Michael Habeck, Marion
  Subklewe, Daniel~N. Wilson, and Roland Beckmann.
\newblock {Structures of the human and Drosophila 80S ribosome}.
\newblock {\em Nature}, 2013.

\bibitem{Avramov2019DeepMaps}
Todor~Kirilov Avramov, Dan Vyenielo, Josue Gomez-Blanco, Swathi Adinarayanan,
  Javier Vargas, and Dong Si.
\newblock {Deep learning for validating and estimating resolution of
  cryo-electron microscopy density maps}.
\newblock {\em Molecules}, 2019.

\bibitem{Bendory2019Single-particleOpportunities}
Tamir Bendory, Alberto Bartesaghi, and Amit Singer.
\newblock {Single-particle cryo-electron microscopy: Mathematical theory,
  computational challenges, and opportunities}.
\newblock 8 2019.

\bibitem{Bepler2018Positive-unlabeledMicrographs}
Tristan Bepler, Andrew Morin, Alex~J. Noble, Julia Brasch, Lawrence Shapiro,
  and Bonnie Berger.
\newblock {Positive-unlabeled convolutional neural networks for particle
  picking in cryo-electron micrographs}.
\newblock {\em Lecture Notes in Computer Science (including subseries Lecture
  Notes in Artificial Intelligence and Lecture Notes in Bioinformatics)}, 10812
  LNBI:245--247, 2018.

\bibitem{Bepler2019ExplicitlySpatial-VAE}
Tristan Bepler, Ellen~D. Zhong, Kotaro Kelley, Edward Brignole, and Bonnie
  Berger.
\newblock {Explicitly disentangling image content from translation and rotation
  with spatial-VAE}.
\newblock 9 2019.

\bibitem{Bergstra2012RandomOptimization}
James Bergstra and Yoshua Bengio.
\newblock {Random search for hyper-parameter optimization}.
\newblock {\em Journal of Machine Learning Research}, 13:281--305, 2012.

\bibitem{Boesen2016AutoencodingMetric}
Anders Boesen, Lindbo Larsen, Søren~Kaae S{\o}nderby, Hugo Larochelle, Ole
  Winther, and Olwi@dtu Dk.
\newblock {Autoencoding beyond pixels using a learned similarity metric}.
\newblock Technical report, 2016.

\bibitem{Chen2018IsolatingAutoencoders}
Ricky T.~Q. Chen, Xuechen Li, Roger Grosse, and David Duvenaud.
\newblock {Isolating Sources of Disentanglement in Variational Autoencoders}.
\newblock (NeurIPS), 2018.

\bibitem{Chen2016InfoGAN:Nets}
Xi Chen, Yan Duan, Rein Houthooft, John Schulman, Ilya Sutskever, and Pieter
  Abbeel.
\newblock {InfoGAN: Interpretable representation learning by information
  maximizing generative adversarial nets}.
\newblock {\em Advances in Neural Information Processing Systems}, pages
  2180--2188, 2016.

\bibitem{Dempster1977Algorithm}
A.~P. Dempster, N.~M. Laird, and D.~B. Rubin.
\newblock { Maximum Likelihood from Incomplete Data Via the EM Algorithm }.
\newblock {\em Journal of the Royal Statistical Society: Series B
  (Methodological)}, 1977.

\bibitem{Detlefsen2019ExplicitModels}
Nicki~Skafte Detlefsen and Søren Hauberg.
\newblock {Explicit Disentanglement of Appearance and Perspective in Generative
  Models}.
\newblock 6 2019.

\bibitem{Fernandez-Leiro2017ARELION}
Rafael Fernandez-Leiro and Sjors~H.W. Scheres.
\newblock {A pipeline approach to single-particle processing in RELION}.
\newblock In {\em Acta Crystallographica Section D: Structural Biology},
  volume~73, pages 496--502. International Union of Crystallography, 6 2017.

\bibitem{Frank1996Two-DimensionalTechniques}
Joachim Frank.
\newblock {Two-Dimensional Averaging Techniques}.
\newblock In {\em Three-Dimensional Electron Microscopy of Macromolecular
  Assemblies}. 1996.

\bibitem{Glorot2010UnderstandingNetworks}
Xavier Glorot and Yoshua Bengio.
\newblock {Understanding the difficulty of training deep feedforward neural
  networks}.
\newblock In {\em Proc. of the Thirteenth International Conference on
  Artificial Intelligence and Statistics, PMLR}, pages 249--256, 2010.

\bibitem{Gnanadesikan2014RobustData}
Ramanathan Gnanadesikan and Jon~R. Kettenring.
\newblock {Robust Estimates , Residuals , and Outlier Detection with
  Multiresponse Data}.
\newblock {\em Biometrics}, 28(1):81--124, 2014.

\bibitem{Goodfellow2014GenerativeNets}
Ian~J. Goodfellow, Jean Pouget-Abadie, Mehdi Mirza, Bing Xu, David
  Warde-Farley, Sherjil Ozair, Aaron Courville, and Yoshua Bengio.
\newblock {Generative Adversarial Nets}.
\newblock Technical report, 2014.

\bibitem{Grant2018CisTEMProcessing}
Timothy Grant, Alexis Rohou, and Nikolaus Grigorieff.
\newblock {CisTEM, user-friendly software for single-particle image
  processing}.
\newblock {\em eLife}, 2018.

\bibitem{Huckemann2010IntrinsicActions}
Stephan Huckemann, Thomas Hotz, and Axel Munk.
\newblock {Intrinsic shape analysis: Geodesic PCA for riemannian manifolds
  modulo isometric lie group actions}.
\newblock {\em Statistica Sinica}, 20(1):1--58, 2010.

\bibitem{Ioffe2015BatchShift}
Sergey Ioffe and Christian Szegedy.
\newblock {Batch Normalization: Accelerating Deep Network Training by Reducing
  Internal Covariate Shift}.
\newblock 2015.

\bibitem{Jahanbegloo2012IdentifyingOutliers}
Ramin Jahanbegloo and Ramin Jahanbegloo.
\newblock {Identifying Density-Based Local Outliers}.
\newblock (October 2017):1--22, 2012.

\bibitem{Kim2018DisentanglingFactorising}
Hyunjik Kim and Andriy Mnih.
\newblock {Disentangling by factorising}.
\newblock {\em 35th International Conference on Machine Learning, ICML 2018},
  6:4153--4171, 2018.

\bibitem{Kimanius2016AcceleratedRELION-2}
Dari Kimanius, Björn~O. Forsberg, Sjors~H.W. Scheres, and Erik Lindahl.
\newblock {Accelerated cryo-EM structure determination with parallelisation
  using GPUS in RELION-2}.
\newblock {\em eLife}, 2016.

\bibitem{Kingma2015Adam:Optimization}
Diederik~P. Kingma and Jimmy Lei~Ba.
\newblock {Adam: A Method for Stochastic Optimization}.
\newblock In {\em Proc of the 3rd International Conference for Learning
  Representations}, 2015.

\bibitem{Kingma2014Auto-EncodingBayes}
Diederik~P. Kingma and Max Welling.
\newblock {Auto-Encoding Variational Bayes}.
\newblock In {\em Proceedings of the 2nd International Conference on Learning
  Representations (ICLR)}, 2014.

\bibitem{Li2018MassivelyTuning}
Liam Li, Kevin Jamieson, Afshin Rostamizadeh, Ekaterina Gonina, Moritz Hardt,
  Benjamin Recht, and Ameet Talwalkar.
\newblock {Massively Parallel Hyperparameter Tuning}.
\newblock pages 1--16, 2018.

\bibitem{Liu2008IsolationForest:Forest}
Fei~Tony Liu, Kai~Ming Ting, and Zhi-Hua Zhou.
\newblock {IsolationForest: Isolation Forest}.
\newblock {\em 2008 Eigth IEEE International Conference on Data Mining}, 2008.

\bibitem{Lopez2018InformationBayes}
Romain Lopez, Jeffrey Regier, Michael~I. Jordan, and Nir Yosef.
\newblock {Information constraints on auto-encoding variational Bayes}.
\newblock {\em Advances in Neural Information Processing Systems},
  2018-Decem(1):6114--6125, 2018.

\bibitem{Louizos2015TheAutoencoder}
Christos Louizos, Kevin Swersky, Yujia Li, Max Welling, and Richard Zemel.
\newblock {The Variational Fair Autoencoder}.
\newblock pages 1--11, 2015.

\bibitem{Nogales2015TheTechnique}
Eva Nogales.
\newblock {The development of cryo-EM into a mainstream structural biology
  technique}.
\newblock {\em Nature Methods}, 2015.

\bibitem{Postnikov2001}
Mikhail Postnikov.
\newblock {\em {Riemannian Geometry}}.
\newblock Encyclopaedia of Mathem. Sciences. Springer, 2001.

\bibitem{Punjani2017CryoSPARC:Determination}
Ali Punjani, John~L. Rubinstein, David~J. Fleet, and Marcus~A. Brubaker.
\newblock {CryoSPARC: Algorithms for rapid unsupervised cryo-EM structure
  determination}.
\newblock {\em Nature Methods}, 2017.

\bibitem{Rezende2014StochasticModels}
Danilo~J. Rezende, Shakir Mohamed, and Daan Wierstra.
\newblock {Stochastic Backpropagation and Approximate Inference in Deep
  Generative Models}.
\newblock In {\em Proceedings of the 31st International Conference on Machine
  Learning}, 2014.

\bibitem{Rohou2015CTFFIND4:Micrographs}
Alexis Rohou and Nikolaus Grigorieff.
\newblock {CTFFIND4: Fast and accurate defocus estimation from electron
  micrographs}.
\newblock {\em Journal of Structural Biology}, 2015.

\bibitem{Rullgard2011SimulationSpecimens}
Hans Rullg{\aa}rd, L.-G. {\"{O}}fverstedt, Sergey Masich, Bertil Daneholt, and
  Ozan {\"{O}}ktem.
\newblock {Simulation of transmission electron microscope images of biological
  specimens}.
\newblock {\em Journal of Microscopy}, 2011.

\bibitem{Sanchez-Garcia2018DeepMicroscopy}
Ruben Sanchez-Garcia, Joan Segura, David Maluenda, Jose~Maria Carazo, and
  Carlos Oscar~S. Sorzano.
\newblock {Deep Consensus, a deep learning-based approach for particle pruning
  in cryo-electron microscopy}.
\newblock {\em IUCrJ}, 5:854--865, 2018.

\bibitem{Scheres2012ADetermination}
Sjors~H.W. Scheres.
\newblock {A bayesian view on cryo-EM structure determination}.
\newblock {\em Journal of Molecular Biology}, 2012.

\bibitem{Scheres2015Semi-automatedRELION-1.3}
Sjors~H.W. Scheres.
\newblock {Semi-automated selection of cryo-EM particles in RELION-1.3}.
\newblock {\em Journal of Structural Biology}, 2015.

\bibitem{Thiagarajan2017-VAE:Framework}
Ganesh Thiagarajan and George~Z. Voyiadjis.
\newblock {{$\beta$}-VAE: Learning basic visual concepts with a constrained
  variational framework}.
\newblock {\em Proc. of ICLR}, pages 1--13, 2017.

\bibitem{Thompson2019CollectionMicroscopy}
Rebecca~F. Thompson, Matthew~G. Iadanza, Emma~L. Hesketh, Shaun Rawson, and
  Neil~A. Ranson.
\newblock {Collection, pre-processing and on-the-fly analysis of data for
  high-resolution, single-particle cryo-electron microscopy}.
\newblock {\em Nature Protocols}, 2019.

\bibitem{Wagner2019SPHIRE-crYOLOCryo-EM}
Thorsten Wagner, Felipe Merino, Markus Stabrin, Toshio Moriya, Claudia Antoni,
  Amir Apelbaum, Philine Hagel, Oleg Sitsel, Tobias Raisch, Daniel Prumbaum,
  Dennis Quentin, Daniel Roderer, Sebastian Tacke, Birte Siebolds, Evelyn
  Schubert, Tanvir~R. Shaikh, Pascal Lill, Christos Gatsogiannis, and Stefan
  Raunser.
\newblock {SPHIRE-crYOLO is a fast and accurate fully automated particle picker
  for cryo-EM}.
\newblock {\em Communications Biology}, 2(1):1--13, 2019.

\bibitem{Wang2016DeepPicker:Cryo-EM}
Feng Wang, Huichao Gong, Gaochao Liu, Meijing Li, Chuangye Yan, Tian Xia,
  Xueming Li, and Jianyang Zeng.
\newblock {DeepPicker: A deep learning approach for fully automated particle
  picking in cryo-EM}.
\newblock {\em Journal of Structural Biology}, 195(3):325--336, 2016.

\bibitem{Zheng2017MotionCor2:Microscopy}
Shawn~Q. Zheng, Eugene Palovcak, Jean~Paul Armache, Kliment~A. Verba, Yifan
  Cheng, and David~A. Agard.
\newblock {MotionCor2: Anisotropic correction of beam-induced motion for
  improved cryo-electron microscopy}, 2017.

\bibitem{Zhong2019ReconstructingModels}
Ellen~D. Zhong, Tristan Bepler, Joseph~H. Davis, and Bonnie Berger.
\newblock {Reconstructing continuously heterogeneous structures from single
  particle cryo-EM with deep generative models}.
\newblock pages 1--16, 2019.

\bibitem{Zhu2017AMicroscopy}
Yanan Zhu, Qi Ouyang, and Youdong Mao.
\newblock {A deep convolutional neural network approach to single-particle
  recognition in cryo-electron microscopy}.
\newblock {\em BMC Bioinformatics}, 18(1):1--10, 2017.

\bibitem{Zivanov2018NewRELION-3}
Jasenko Zivanov, Takanori Nakane, Björn~O. Forsberg, Dari Kimanius, Wim~J.H.
  Hagen, Erik Lindahl, and Sjors~H.W. Scheres.
\newblock {New tools for automated high-resolution cryo-EM structure
  determination in RELION-3}.
\newblock {\em eLife}, 2018.

\end{thebibliography}

}

\end{document}